\documentstyle[12pt]{article}
\newcommand{\Tc}{T_{c} }
\newcommand{\Fc}{F_{\rm cb} }
\newcommand{\Rc}{R_{\rm cb} }

\newcommand{\Fs}{F_{\rm sb} }
\newcommand{\ns}{n_{\rm sb} }

\begin{document}

\begin{flushright}
FERMILAB-Pub-95/304-A \\
DART-HEP-95/05\\
Sept. 1995
\end{flushright}

\vspace{1in}
\begin{center}
{\Large{\bf  Nonperturbative effects on nucleation}}\\
\vspace{.4in}

Marcelo Gleiser \\
{\em Department of Physics and Astronomy, Dartmouth College,}\\
{\em Hanover, NH 03755, USA }\\{\em email: gleiser@peterpan.dartmouth.edu}
\vspace{.15in}

Andrew F. Heckler \\
{\em  NASA/Fermilab Astrophysics Center,
Fermi National Accelerator Laboratory,}\\
{\em Batavia, IL 60510, USA}\\{\em email: aheckler@fnas04.fnal.gov}

\end{center}

\begin{abstract}
A nonperturbative correction to the thermal nucleation
rate of critical bubbles in a first order phase transition
 is estimated. The correction originates
from large-amplitude fluctuations which may be present
before the transition occurs. Using a simple model of a scalar field
in a double-well potential, we present a method to
obtain a corrected potential which incorporates
the free-energy density available from large-amplitude fluctuations,
which is not included in the usual perturbative calculation.
For weaker phase transitions, the nucleation rate can be
much larger than the rate calculated via perturbation theory.
As an application of our method, we show how nonperturbative
corrections can both qualitatively and quantitatively explain
anomalously high nucleation rates observed in 2-d numerical simulations.
\end{abstract}

\newpage

Although the simplest first order phase transitions are characterized by a
discontinuous jump of a scalar order parameter
between two distinct phases, they do not all proceed in the
same way \cite{REVIEW}. For very strong first order phase
transitions, where the free-energy barrier between the phases
is large, the transition is initiated by the nucleation of critical-sized
bubbles of the new phase in the background of the  metastable ({\it e.g.},
super-cooled) old phase. By definition, these critical bubbles are just
large enough to overcome their surface tension and grow, eventually
converting the whole medium to the new phase.
The large barrier between
the two phases suppresses
large-amplitude thermal fluctuations of the order parameter; an initial
metastable state is well-defined, as
no fraction of the volume is in the new phase before the transition
occurs.  In this case, the metastable phase can be regarded as
``homogeneous'', as only very small-amplitude thermal fluctuations are
present.
This is the situation described by Langer's theory of homogeneous
nucleation \cite{LANGER}, or, in the context of relativistic quantum field
theories, by the work of Coleman and Callan \cite{COLEMAN}.

Besides the decay of the ``near-homogeneous'' metastable state
described by nucleation theory, one can investigate the evolution
of an unstable initial state
which is characterized by considerable phase mixing.
Within the context of
condensed matter systems, this situation corresponds to a quench within
the unstable ``spinodal'' region of the two-phase diagram. In this case,
the two phases separate by
the mechanism known as ``spinodal decomposition'';
small-amplitude, long-wavelength fluctuations grow exponentially fast,
forming domains of the two phases which will eventually coarsen,
as the system approaches its final equilibrium state.

In this letter we will address the dynamics of phase transitions
characterized by an initial state which lies within the ``grey zone''
between homogeneous nucleation and spinodal decomposition.
Looking at the whole ``spectrum'' of first order phase transitions,
from very strong to very weak, it is clear that the
amount of phase-mixing of the initial state will strongly influence
the subsequent dynamics of the transition. However, the standard
method of calculating the nucleation rate employs Gaussian
perturbation theory, which
is valid only for small amplitude
fluctuations \cite{GAUSSIAN}.
For strong transitions this
approximation is valid. But for weaker transitions, large amplitude
fluctuations are more abundant, and can have an important effect. Our
goal is to present an approximate
 method by which the presence of large-amplitude
fluctuations is consistently incorporated into the calculation of nucleation
rates. Thus, we are implicitly assuming that we are close enough to
the regime described by homogeneous nucleation that we can still distinguish
between the two low-temperature phases.

Large-amplitude thermal fluctuations will be modelled
by the so-called sub-critical bubble
method \cite{GKW}. Recent results \cite{GHK} have shown that modelling the
dominant fluctuations by sub-critical bubbles is in excellent agreement with
3-d simulations \cite{BG}.
The model utilizes the fact that along with the nucleation of critical bubbles
in the meta-stable phase,
smaller size, though still large amplitude, ``sub-critical'' bubbles will
also be nucleated (and in much greater number because they have a lower
free energy). These bubbles by definition will always shrink and eventually
disappear, but there will always be some non-zero equilibrium
number density $n_{\rm sb}$
at a given temperature. Their presence may lead to large corrections on
nucleation rates.

To begin, let us consider the standard model of a phase transition,
in which the order parameter  is a real scalar field $\phi$, which has a
quartic double-well potential of the form
\begin{eqnarray}\label{V}
V(\phi) = \frac{1}{2} m^2 \phi^2 - \frac{1}{6}g \phi^{3 } +\frac{h}{24} \phi^4.
\end{eqnarray}
This potential has two minima, one at $\phi = 0$ and
at $\phi = \phi_{+}$, which represent the two phases of the system.
It can be thought of as the homogeneous part of a typical phenomenological
Ginzburg-Landau coarse-grained free-energy density (the cubic term can
always be made into a linear term), or as some effective potential where
additional degrees of freedom coupled to $\phi$ have been integrated out.
Our analysis will be purely classical, valid for
$T\gg m$, where $m$ is the mass
of the low-energy mesonic
excitations in the associated quantum theory.
All relevant field configurations contain many quanta.

We would like to incorporate the
free-energy density associated with large-amplitude,
nonperturbative fluctuations into the computation of the decay rate. In the
spirit of the renormalization group approach, this should be equivalent to
an effective
``coarse-graining'' of the classical
potential; averaging over these large-amplitude fluctuations will lead to a
shift in the background free-energy density and decay barrier, which in
principle can be translated into
a change in the bare couplings of the model.
We can understand how
to estimate the effective coarse-graining by first studying the thin-wall
limit of critical bubble nucleation.

In the standard theory,
which neglects phase mixing, the nucleation rate $\Gamma$  is
proportional to $e^{-F_{\rm cb}/T}$, where $F_{\rm cb}$
is free energy needed to form
a critical bubble in the metastable
background. For an arbitrary thin-walled spherical
bubble of radius $R$ and amplitude $\phi_{\rm thin}{\
\lower-1.2pt\vbox{\hbox{\rlap{$<$}\lower5pt\vbox{\hbox{$\sim$}}}}\ } \phi_+$,
where thin-walled
means the radius $R$ is much greater than the bubble wall thickness,
the free energy of the bubble takes the well-known form \cite{LINDE}
\begin{eqnarray}\label{Fthin}
F_{\rm thin}(R) &=&  4\pi R^2 \sigma - \frac{4 \pi}{3}R^3 \Delta V  .
\end{eqnarray}
This formula has a simple physical interpretation. The first term
is the energy it costs to form the bubble wall, where
$\sigma\equiv \frac{1}{2}\int dr(\partial \phi/\partial r)^2$ is the
surface tension. The second term is the energy ``gained'' by
converting a spherical volume of the metastable phase into
the lower energy phase.
Therefore,  $\Delta V$ is defined as the difference in
free-energy density between the background medium and the
bubble's interior.
Since $\phi_{\rm thin} {\
\lower-1.2pt\vbox{\hbox{\rlap{$<$}\lower5pt\vbox{\hbox{$\sim$}}}}\ } \phi_{+}$,
for a homogeneous
background (metastable) we can write,
\begin{eqnarray}\label{}
\Delta V_{0} = V(0) - V(\phi_{+}),
\end{eqnarray}
where we have explicitly used the subscript $0$ to stress
that this is for the case with no phase mixing.

If there is significant phase mixing in the background metastable state,
its free-energy density is no longer $V(0)$.
One must also account for the free-energy density of the nonperturbative,
large-amplitude fluctuations. Since there is no formal way of deriving
this contribution outside improved perturbative schemes, we
propose to estimate the corrections to the background free-energy
density by following another route. We start by writing
\begin{eqnarray}\label{fullfree}
{\rm free-energy\, density\, of\, metastable\,\, state} = V(0) +
{\cal F}_{\rm sc},
\end{eqnarray}
where ${\cal F}_{\rm sc}$ is the nonperturbative contribution to
the free-energy density due to the large amplitude fluctuations, which we
assume can be modelled by
subcritical bubbles. We will calculate  ${\cal F}_{\rm sc}$
further below.

We thus define the effective free-energy difference
$\Delta V_{\rm cg}$, which includes
corrections due to phase mixing, as
\begin{eqnarray}\label{DeltaV}
\Delta V_{\rm cg}= \Delta V_{0} + {\cal F}_{\rm sc}
\end{eqnarray}
which is the sum of the free-energy difference calculated in
the standard way [eq.~(\ref{Fthin})],
and the ``extra'' free-energy density
due to the presence of subcritical bubbles. Henceforth, the subscript `cg'
will stand for ``coarse-grained''.

We note that while we have made a correction to
$\Delta V$, we have not made any correction to the surface
tension $\sigma$. Since we are considering the thin-wall limit,
as long as $\left\langle  \phi \right\rangle$ is small, which is
true if subcritical bubbles do not occupy a large fraction of space,
the correction to $\sigma$ will be subdominant.
[Note that the presence of subcritical bubbles may shift $\left\langle  \phi
\right\rangle$ by roughly $\sum_ie^{-F_i/T}\phi_i$, where $\phi_i$ is the
amplitude of a given fluctuation, and $F_i$ its associated free energy.]
Thus, the arguments here give
a lower bound on the magnitude of the corrections. Later on, both volume and
surface corrections will be automatically included in the calculation.

Since a critical size bubble is defined as the bubble for which all
forces on the bubble wall cancel, {\it i.e.} $\partial F/\partial R|_{\Rc}=0$,
we can now use eq.~(\ref{Fthin}) to obtain the free energy needed to
form a thin-wall critical bubble in a background with subcritical bubbles
\begin{eqnarray}\label{Fcritthin}
\Fc = \frac{2 \pi}{3} \Rc^3 (\Delta V_{0} + {\cal F}_{\rm sc}) =
\frac{16 \pi}{3} \frac{\sigma^3}{ (\Delta V_{0} + {\cal F}_{\rm sc})^{2}}
\end{eqnarray}
and the radius of the critical bubble is
\begin{eqnarray}\label{Rcritthin}
\Rc = \frac{2 \sigma}{\Delta V_{0} + {\cal F}_{\rm sc}}.
\end{eqnarray}
Equations~(\ref{Fcritthin}) and (\ref{Rcritthin}) warrant several
comments. First, in the limit of a very strong phase transition,
subcritical
bubbles are suppressed (${\cal F}_{\rm sc}\rightarrow 0$), and
both $\Fc$ and $\Rc$ approach the standard expressions for the
free energy and radius of a critical bubble in a homogeneous background.
Second, notice that ${\cal F}_{\rm sc}$ acts in the same way as
the free-energy difference $\Delta V_{0}$. The
presence of subcritical bubbles is equivalent to extra free energy in the
medium, which enhances the nucleation of critical bubbles.
In particular, for potentials near degeneracy such that
$\Delta V_{0} {\
\lower-1.2pt\vbox{\hbox{\rlap{$<$}\lower5pt\vbox{\hbox{$\sim$}}}}\ } {\cal
F}_{\rm sc}$, the nucleation rate of
critical bubbles  $\Gamma \sim e^{-\Fc/T}$, can be {\em much} greater
than in the case ignoring the presence of subcritical bubbles.

Finally, notice that as $\Delta V_{0} \rightarrow 0$, neither
the critical-bubble energy nor its radius become infinite. For
temperature-dependent potentials which (ignoring the corrections
from subcritical bubbles) are degenerate
at the critical temperature $\Tc$,
the nucleation rate
$\Gamma \sim e^{-\Fc/\Tc}$ is finite. In fact, the nucleation rate
of critical bubbles may be non-zero even {\em above} the critical
temperature (again, using the uncorrected expression for the potential).
This is a
testable prediction of our method which, of course, is sensitive to the
equilibrium number-density of subcritical bubbles.

This final comment suggests an important point. Since for degenerate
potentials  (temperature dependent or not)
no critical bubbles should be nucleated, taking into
account subcritical bubbles must lead to a change in the coarse-grained
free-energy density (or potential)
describing the transition. Thus, it should be possible to translate the
``extra'' free energy available in the system due to the presence of
subcritical bubbles in the background into a corrected potential for the
scalar order parameter. We will write this corrected potential as
$ V_{\rm cg}(\phi)$.

The standard coarse-grained free energy is calculated by integrating out
the short wavelength modes (usually up to the correlation length)
from the partition function
of the system, and is approximated by
the familiar form \cite{Langer74a}
\begin{eqnarray}\label{F}
F_{\rm cg} = \int d^3r \left( \frac{1}{2}(\nabla \phi)^2 +
V_{\rm cg}(\phi)\right)~.
\end{eqnarray}

How do we estimate $V_{\rm cg}$? One way
is to simply constrain it to be consistent with the thin wall limit.
That is, as $V_{\rm cg}(\phi)$ approaches degeneracy
({\it i.e.} $\Delta V_{\rm cg}(\phi)\rightarrow 0$), it must obey the
thin wall limit of eq.~(\ref{DeltaV}). Note that with a simple rescaling,
the potential of eq.~(\ref{V}) can be written in terms of one free parameter.
Thus, the thin
wall constraint can be used to express the corrected value of
this parameter in terms of ${\cal F}_{\rm sc}$ in appropriate units.
The free energy of the critical bubble is then obtained
by finding the bounce solution to the equation of motion
$\nabla^{2}\phi - dV_{\rm cg}(\phi)/d\phi=0$ by the usual shooting method,
and substituting this solution
into  eq.~(\ref{F}).

Therefore, in order to determine $V_{\rm cg}$, we must first calculate the
free-energy density ${\cal F}_{\rm sc}$ of the subcritical bubbles.
As a first step, we follow  the work of  Ref.~\cite{GHK},
to obtain the equilibrium number density
$n_{\rm sb}$ of subcritical bubbles. If we define the distribution function
$f\equiv \partial^{2}n_{\rm sb}/\partial R\partial \phi_{A}$, then
$f(R,\phi_{A},t)dR d\phi_{A}$ is the number density of bubbles
with radius between $R$ and $R+dR$ and amplitude between
$\phi_{A}$ and $\phi_{A} + d\phi_{A}$ at time $t$. It satisfies the
Boltzmann equation,
\begin{eqnarray}\label{boltz}
\frac{\partial f(R,\phi_{A},t)}{\partial t} &=&
-|v|\frac{\partial f}{\partial R} + (1-\gamma )G_{0\rightarrow +} \nonumber \\
& & - f{\cal V}G_{\rm Therm} - \gamma G_{+\rightarrow 0}.
\end{eqnarray}
The first term on the RHS is the shrinking term (note that
$ v= \partial R/\partial t$ is negative), the second term is the
nucleation term where $G$ is the nucleation distribution function,
which is defined by $\Gamma = \int{dRd\phi G}$, and
$\Gamma_{0\rightarrow +}$ is the nucleation rate per unit volume of
subcritical bubbles from the ``0'' phase (the initial phase) to
the ``+'' phase. The division of the system into two phases depends on the
particular application at hand, as will be clear in the example below.
By the Gibb's distribution,
$G_{0\rightarrow +} = Ae^{-\Fs(R,\phi_{A})/T}$, where $A$ is a
constant independent of $R$ and $\phi$.

The factor $\gamma$ is defined
as the fraction of volume in the ``+'' phase, and is obtained by summing over
subcritical bubbles of all amplitudes within this phase. The third term is a
phenomenological thermal destruction term (see work by Gelmini and Gleiser
in Ref.~\cite{GKW}),
where ${\cal V}$ is the volume of a bubble of radius $R$, and
$G_{\rm Therm} = aT/{\cal V}$, where $a$ is a constant. The fourth
term is the inverse nucleation term. For more details about this
Boltzmann equation, see Ref.~\cite{GHK}, which has improved upon
the work of Gelmini and Gleiser (Ref.~\cite{GKW}).

The free energy of the subcritical bubbles is determined by
modelling them as Gaussian fluctuations
with amplitude $\phi_{A}$ and radius $R$,
\begin{eqnarray}\label{}
\phi_{\rm sc}(r) = \phi_{A} e^{-r^2/R^2}.
\end{eqnarray}
The free energy of a given configuration can then be found by using
the general formula,
\begin{eqnarray}\label{Fs}
\Fs = \int d^3r \left( \frac{1}{2}(\nabla \phi_{\rm sc})^2 +
V(\phi_{\rm sc})\right).
\end{eqnarray}
Although this approach only includes one particular shape out of all possible
field configurations, the agreement between theory and numerical experiments
indicates that the Gaussian profile is an adequate {\it ansatz} for the
dominant large-amplitude thermal fluctuations.

The equilibrium number density of subcritical bubbles is
found by solving eq.~(\ref{boltz}) with $\partial f/\partial t =0$,
imposing the physical boundary condition $f(r\rightarrow\infty)=0$.
Once we know the distribution function and free energy
for a bubble of a given radius $R$ and amplitude $\phi_{A}$,
we can estimate the total energy density of the Gaussian
subcritical bubbles, summed over all relevant radii and amplitudes.
We can write, in general,
\begin{eqnarray}\label{calF}
{\cal F}_{\rm sc} \approx  \int_{\phi_{\rm min}}^{\infty}
{\int_{R_{\rm min}}^{R_{\rm max}}{F_{\rm sb}
\frac{\partial^{2}\ns}{\partial R \partial \phi_{A}}dRd\phi_{A}}},
\end{eqnarray}
where $\phi_{\rm min}$ defines the lowest amplitude within the ``+'' phase,
typically (but not necessarily)
taken to be the maximum of the double-well potential. $R_{\rm min}$ is the
smallest radius for the subcritical bubbles, compatible with the
coarse-graining scale. For example, it can be
a lattice cut-off in numerical simulations,
or the mean-field correlation length in continuum models. As for $R_{\rm max}$,
it is natural to choose it to be the critical bubble radius.

As an application of the above method, we will investigate nucleation rates
in the context of a 2-d model for which accurate numerical results
are available \cite{Alford93a}.
This will allow us to compare the results obtained by incorporating
subcritical bubbles into the calculation of the decay barrier with the
results from the numerical simulations.

The 2-d scalar potential $V(\phi)$ is given in
eq.~(\ref{V}).
Following the rescaling of Ref.~ \cite{Alford93a},
the potential can be written in terms of one dimensionless
parameter $\lambda\equiv
m^2h/g^2$,
\begin{eqnarray}\label{V2d}
V(\phi) = \frac{1}{2} \phi^2 - \frac{1}{6} \phi^{3 }
+\frac{\lambda}{24} \phi^4.
\end{eqnarray}
This double-well potential is degenerate when $\lambda = 1/3$,
and the second minimum is lower than the first when $\lambda <1/3$.

As argued before,
we find the new coarse-grained potential $V_{\rm cg}$
(or, equivalently, $\lambda_{\rm cg}$) by constraining it to agree with the
thin wall limit. Simple algebra from eqs.~(\ref{DeltaV}) and (\ref{V2d})
yields, to first order in the deviation from degeneracy,
\begin{eqnarray}\label{}
\lambda_{\rm cg} = \lambda - \frac{{\tilde {\cal F}}_{\rm sc}}{54}
\end{eqnarray}
where ${\tilde {\cal F}}_{\rm sc}=\frac{g^2}{m^6}{\cal F}_{\rm sc}$ is
the dimensionless free-energy density in subcritical bubbles.
The new potential $V_{\rm cg}$ is then used to
find the bounce solution and the free energy of the critical bubble.

The calculation of ${\cal F}_{\rm sc}$ in two dimensions is fairly
straightforward. Close to the thin wall limit
({\it i.e.}, $ G_{0\rightarrow +}\approx
G_{+\rightarrow 0}\equiv G$), one can analytically solve the
equilibrium Boltzmann equation for the density distribution function,
obtaining
$f(R,\phi_{A},T)=(1-2\gamma)W_T(R,\phi_{A})$, where $(v\equiv |v|)$
\begin{eqnarray}
W_T(R,\phi_{A}) & = & \frac{(A/v)}{2}
{\rm exp}\left [
-\frac{\alpha}{T}+RT(a/v)+\frac{(a/v)^2T^3}{4\beta}\right ] \nonumber \\
& & \sqrt{\frac{\pi T}{\beta}}
\left \{1-{\rm erf}\left [\sqrt{\frac{\beta}{T}}\left (
R+\frac{(a/v)T^2}{2\beta^2}
\right ) \right] \right \}~,
\end{eqnarray}
and we wrote the free energy of a given subcritical configuration as
$F_{\rm sb}\equiv \alpha+\beta R^2$, with $\alpha=\pi\phi_A^2/2$, and
$\beta=\alpha\left (\frac{1}{2}-\frac{1}{9}\phi_A+\frac{\lambda}{48}\phi_A^2
\right )$.
The fraction of the volume occupied by subcritical bubbles is then,
\begin{equation}
\gamma(\phi_{\rm max},R_{\rm min},R_{\rm max})=
\frac{I_T(\phi_{\rm max},R_{\rm min},R_{\rm max})}{\left [1+2
I_T(\phi_{\rm max},R_{\rm min},R_{\rm max})\right ]}~,
\end{equation}
where,
$I_T=\int_{\phi_{\rm max}}^{\infty}
\int_{R_{\rm min}}^{R_{\rm max}}\pi r^2W_Tdrd\phi~$.
The radial integration can be done analytically, although the result is not
particularly illuminating. The integral over amplitudes must be done
numerically. We then substitute $f(R,\phi_{A},T)$ and $\gamma$
into eq. (\ref{calF}) to finally find
${\cal F}_{\rm sc}(\lambda,T,A/v,a/v)$.

In order to illustrate the effect of subcritical bubbles on the nucleation
barrier, in Figure 1 we compare the value for the barrier obtained with and
without the corrections, as a function of $\lambda$, for constant values
of the temperature. The temperatures are chosen to be within the range used in
the 2-d simulation. The constant $A$ was fixed at $A=0.02$, consistent with
the measurements of Ref.~\cite{Alford93a}.
Notice that the presence of
subcritical bubbles greatly decreases the barrier as the
potential approaches degeneracy $(\lambda \rightarrow 1/3)$. However, for
small temperatures $T< 10m^4/g^2$, the correction becomes negligible.

In Figure 2 we show that the calculation of the nucleation barrier including
the effects of subcritical bubbles is consistent with data from lattice
simulations,
whereas the standard calculation overestimates the barrier by a large margin.
In fact, the inclusion of subcritical bubbles provides a reasonable
explanation for the anomalously high nucleation rates observed in the
simulations close to degeneracy.
The error bars are from the numerical measurements of the barrier; for larger
values of $\lambda$, higher temperatures had to be used to attain
nucleation, increasing the error in the barrier measurements. However,
we note that even with the large error bars, the data is inconsistent with
the theoretical predictions for the barriers, while the corrected barrier
values fall within the error bars for a wide range of parameters.
We
note that data from 1-d simulations also show the same behavior as
the data in
Figure 2 \cite{Alford93a}.
Simulations in 3-d are in progress, and will enable us to test
this method in more detail.

Finally, we note that the inclusion of nonperturbative corrections through
the definition of an effective, ``coarse-grained'', coupling may have
several consequences not only to the nucleation rate of first order
transitions, but also to their dynamics. Clearly, once we have a
corrected potential, quantities such as the critical temperature, the
amount of super-cooling, the bubble-wall velocities, and the completion time
for the transition will change. This opens up several
possible applications of this method, from laboratory studies of nucleation
to cosmological phase transitions.

We thank S. Dodelson, J. Frieman, E.W. Kolb, A. Stebbins, and E.J. Weinberg
for stimulating discussions.
MG was partially supported by the National Science Foundation through a
Presidential
Faculty Fellows Award (PHY-9453431), and by  NASA (NAGW-4270). He thanks
the Nasa/Fermilab Astrophysics Center for their kind hospitality during
part of this work.
AFH was supported in part by the DOE and by NASA (NAG5-2788) at Fermilab.

\listoffigures

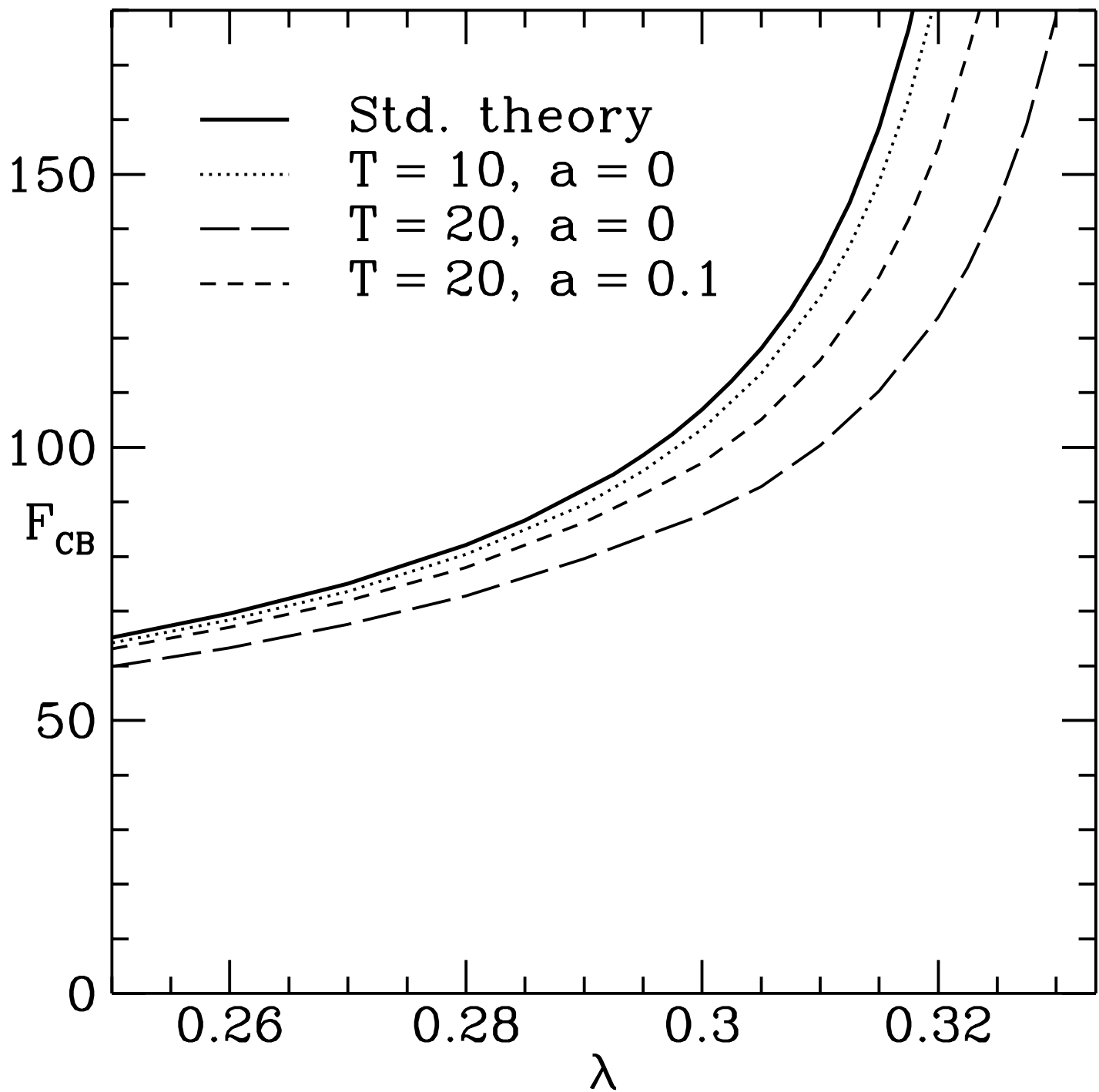
Figure 1. Comparison of the decay barrier as a function of $\lambda$ with and
without the inclusion of subcritical bubbles, at fixed temperatures.\\

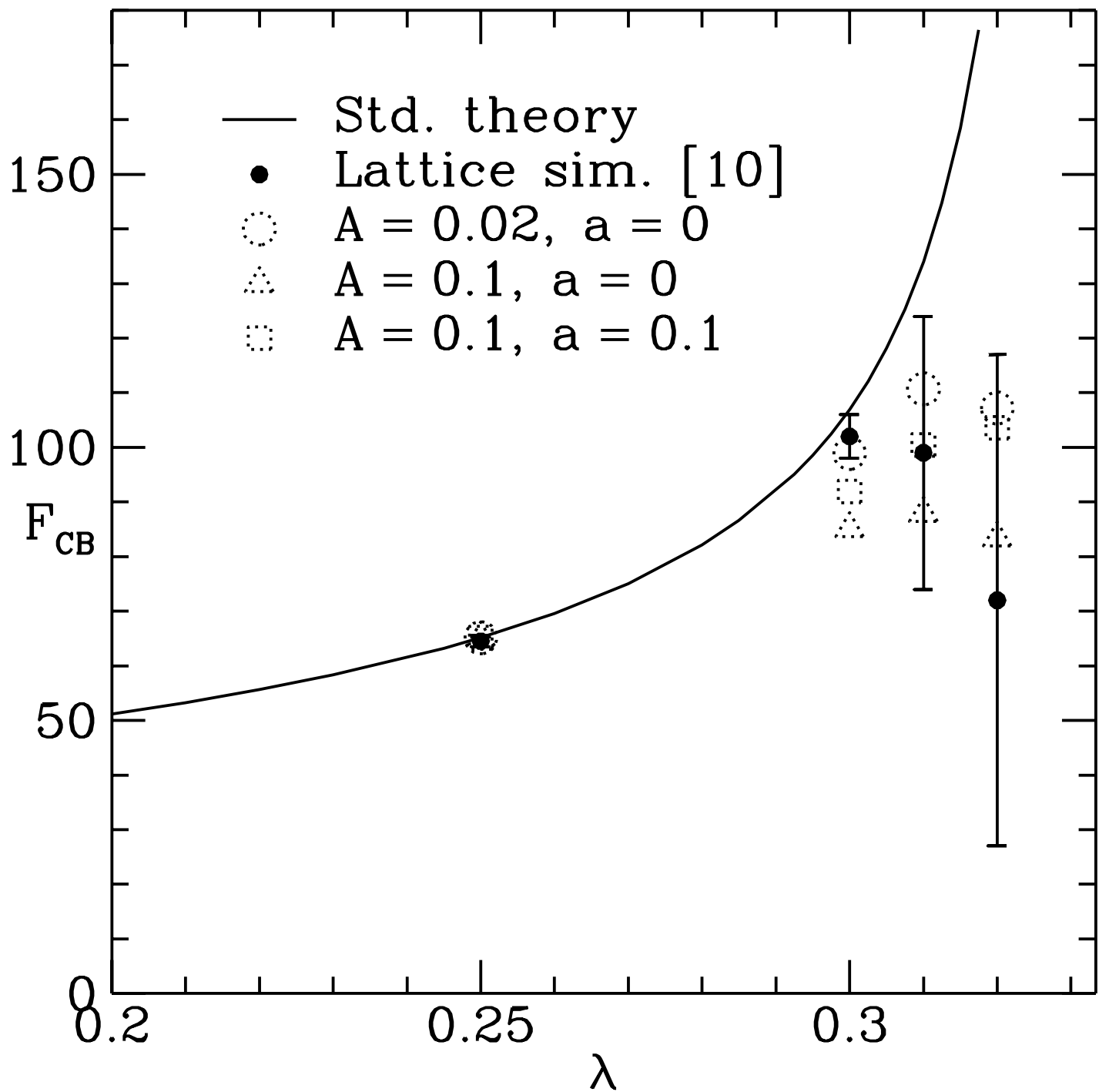
Figure 2. Comparison between numerical data and theoretical predictions
for the decay barrier with and without the inclusion of subcritical
bubbles.\\

\end{document}